\documentclass[aps,showpacs,11pt,superscriptaddress,preprintnumbers,amsmath,amssymb,nofootinbib]{revtex4-1}
\pdfoutput=1
\usepackage{mathrsfs}
\usepackage{epsfig}
\usepackage{graphicx}
\usepackage{dcolumn}
\usepackage{bm}
\usepackage{amsmath}
\usepackage{slashed}       
\usepackage{amssymb}
\usepackage{amsfonts}

\newcommand{\PreserveBackslash}[1]{\let\temp=\\#1\let\\=\temp}
\newcolumntype{C}[1]{>{\PreserveBackslash\centering}p{#1}}
\newcolumntype{R}[1]{>{\PreserveBackslash\raggedleft}p{#1}}
\newcolumntype{L}[1]{>{\PreserveBackslash\raggedright}p{#1}}

\let\jnfont=\rm
\def\NPB#1,{{\jnfont Nucl.\ Phys.\ B }{\bf #1},}
\def\PLB#1,{{\jnfont Phys.\ Lett.\ B }{\bf #1},}
\def\EPJC#1,{{\jnfont Eur.\ Phys.\ Jour.\ C }{\bf #1},}
\def\PRD#1,{{\jnfont Phys.\ Rev.\ D }{\bf #1},}
\def\PRL#1,{{\jnfont Phys.\ Rev.\ Lett.\ }{\bf #1},}
\def\MPLA#1,{{\jnfont Mod.\ Phys.\ Lett.\ A }{\bf #1},}
\def\JPG#1,{{\jnfont J.\ Phys.\ G}{\bf #1},}
\def\CTP#1,{{\jnfont Commun.\ Theor.\ Phys.\ }{\bf #1},}
\def\ZPC#1,{{\jnfont Z.\ Phys.\ C }{\bf #1},}
\def\JHEP#1,{{\jnfont JHEP \ }{\bf #1},}
\def\lsim{\raise0.3ex\hbox{$<$\kern-0.75em\raise-1.1ex\hbox{$\sim$}}}
\def\gsim{\raise0.3ex\hbox{$>$\kern-0.75em\raise-1.1ex\hbox{$\sim$}}}

\newcommand{\GeV}{~\rm GeV}

\newcommand{\fbm}{{~\rm fb}^{-1}}

\begin{document}
\preprint{\parbox{1.2in}{\noindent arXiv: 1601.02570}}

\title{Interpreting the 750 GeV diphoton excess in the Minimal Dilaton Model}
\author{Junjie Cao}
\email{junjiec@itp.ac.cn}
\affiliation{Center for Theoretical Physics, School of Physics and Technology, Wuhan University, Wuhan 430072, China}
\affiliation{Department of Physics, Henan Normal University, Xinxiang 453007, China}
\author{Liangliang Shang}
\email{shlwell1988@gmail.com}
\affiliation{Department of Physics, Henan Normal University, Xinxiang 453007, China}
\author{Wei Su}
\email{weisv@itp.ac.cn}
\affiliation{State Key Laboratory of Theoretical Physics,
      Institute of Theoretical Physics, Academia Sinica, Beijing 100190, China}
\author{Yang Zhang}
\email{zhangyang@itp.ac.cn}
\affiliation{State Key Laboratory of Theoretical Physics,
      Institute of Theoretical Physics, Academia Sinica, Beijing 100190, China}
\author{Jinya Zhu}
\email{zhujy@itp.ac.cn}
\affiliation{Center for Theoretical Physics, School of Physics and Technology, Wuhan University, Wuhan 430072, China}

\begin{abstract}
We try to interpret the 750 GeV diphoton excess in the Minimal Dilaton Model, which extends the SM by
adding one linearized dilaton field and vector-like fermions. We first show by analytic formulae in
this framework that the production rates of the $\gamma \gamma$,
$gg$, $Z\gamma$, $ZZ$, $WW^\ast$, $t\bar{t}$ and  $hh$ signals at the $750 {\rm GeV}$ resonance
are only sensitive to the dilaton-Higgs mixing angle $\theta_S$ and the parameter $\eta \equiv v N_X/f$,
where $f$ is the dilaton decay constant and $N_X$ denotes  the number of the fermions.
Then we scan the two parameters by considering various theoretical and
experimental constraints to find the solutions to the diphoton excess. We conclude that the model can
predict the central value of the diphoton rate without conflicting with any constraints.
The signatures of our explanation at the LHC Run II and the vacuum stability at high energy
scale are also discussed.
\end{abstract}


\maketitle

\section{Introduction}
About four years ago, the hint of a 125 GeV Higgs boson was reported in the diphoton channel by both the ATLAS and CMS collaborations based on about
$5\fbm$ data for each collaboration at the 7-TeV LHC \cite{1202-a-com,1202-c-com}, and this led to the great discovery of the Higgs boson in July 2012 \cite{1207-a-com,1207-c-com}.
Recently another excess in the diphoton channel was reported by the first $3.2\fbm$ data at the 13-TeV LHC \cite{1512-a-750,1512-c-750}. This time
the invariant mass of the signal locates around $750 {\rm GeV}$, and its local and global significances are about $3.6\sigma$ and
$2.3 \sigma$ respectively for the ATLAS analysis, and $2.6 \sigma$ and $2 \sigma$ for the CMS analysis.
Interestingly, although there exists a ostensible inconsistence in the width of the resonance \footnote{Currently with insufficient experimental data,
the ATLAS analysis slightly preferred a wide width of the resonance (about $45 {\rm GeV}$) to a narrow width \cite{1512-a-750}, and by contrast
the CMS analysis favored a narrow width \cite{1512-c-750}. Very recently, an analysis by combining both the ATLAS data and the CMS data
was carried out, and it indicated that the narrow width was preferred \cite{new-analysis}.},
both the analyses favored the diphoton production rate at about $4 {\rm fb}$ in the narrow width approximation.
Such a rate is about $10^4$ times larger than the prediction of the Standard Model (SM) with a
$750 {\rm GeV}$ Higgs boson \cite{1101-h-hb}. Obviously, if this excess is confirmed in near future,
it points undoubtedly to the existence of new physics.

So far more than one hundred theoretical papers have appeared to interpret the excess in new physics models
\cite{Diphoton:UsedRate,general-singlet-extension,DiphotonExcess,DiphotonCorrelation,
1512-COS,Dilaton-1,Dilaton-2,vacuum-1,vacuum-2,vacuum-3,vacuum-4}, and most of them employed
the process $g g \to S \to \gamma \gamma$ with $S$ denoting a scalar particle with mass around $750 \ {\rm GeV}$ to fit the data. From
these studies, one can infer two essential ingredients of the explanations. One is that there must exist other charged and colored particles
to generate by loop effects sufficiently large $S \gamma \gamma$ and $S g g$ interactions. The other is, given the fact that no excess
was observed in the channels such as $ZZ$, $WW^\ast$ and $t\bar{t}$ at the LHC Run I, the particle $S$ is preferred to be
gauge singlet dominated so that the branching ratios of $S \to Z Z$, $W W^\ast$, $t \bar{t}$ are not much larger than
that of $S \to \gamma \gamma$. These requirements guide us in seeking for the explanations of the excess.

In this work, we consider interpreting the diphoton excess in the Minimal Dilaton Model (MDM),
which extends the SM by one gauge singlet field called dilaton \cite{min-dilaton1, min-dilaton2, 1311-MDM}.
Just like the traditional dilaton theories \cite{dilaton-before}, the dilaton in this model
arises from a strong interaction theory with approximate scale invariance at a certain high energy scale. The breakdown
of the invariance then triggers the electroweak symmetry breaking, and during this process,
the dilaton as the pseudo Nambu-Goldstone particle of the broken invariance can be naturally light in
comparison with the high energy scale.
Furthermore, this model assumes that all SM particles except for the Higgs field do not interact with the dynamics  sector,
and consequently the dilaton does not couple directly to the fermions and $W$, $Z$ bosons in the SM.
In this sense, the dilaton is equivalent to an electroweak gauge singlet field.
The model also consists of massive vector-like fermions acting as the lightest particles
in the dynamical sector, to which the dilaton naturally couples in order to recover the scale
invariance: $M \to M e^{-\phi/f}$. As a result, the interactions between the dilaton and the photons/gluons
are induced through loop diagrams of these fermions. These characters enable the MDM as a hopeful theory to
explain the diphoton excess through the dilaton production.
Discussing the capability of the MDM in explaining the excess is the aim of this work.

This paper is organized as follows. We first introduce briefly the MDM in Section II,
and present in Section III some analytical formulae which are used to calculate the diphoton rate.
In Section IV, we discuss the constraints on the model, its capability in explaining the excess,
and also the related phenomenology at the LHC Run II.  For completeness, in
section V we turn to discuss the vacuum stability at high energy scale. Finally, we draw our conclusions in Section VI.

\section{the Minimal Dilaton Model}

As introduced in last section, the MDM extends the SM by adding one gauge singlet field $S$,
which represents a linearized dilaton field, and also vector-like fermions $X_i$. The low energy effective Lagrangian
is then written as \cite{min-dilaton1,min-dilaton2}
\begin{eqnarray}
\mathcal L &=
	&	\mathcal L_\text{SM}
		+ {1\over2}\partial_\mu S\partial^\mu S
		+ \sum_{i=1}^{N_X} \bar{X}_i \left(i \slashed{D} - \frac{M_i}{f}S\right) X_i - V(S,\tilde{H}),  \label{MDM}
\end{eqnarray}
where $\mathcal L_\text{SM}$ is the SM Lagrangian without Higgs potential, $f$ is the decay constant of the dilaton $S$,
$M_i$ is the mass of the fermion $X_i$, and $N_X$ is the number of the vector-like fermions. The scalar potential $V(S,\tilde{H})$
contains terms with explicit breaking of the scale invariance, and its general form is given by
\begin{eqnarray}
V(S,\tilde{H})=\frac{m_S^2}{2} S^2+ {\lambda_S\over 4}S^4 + m_H^2 \left|\tilde{H}\right|^2 + \lambda_H \left|\tilde{H}\right|^4 +
\frac{\lambda_{HS}}{2} S^2\left|\tilde{H}\right|^2,  \label{potential}
\end{eqnarray}
where $m_S$, $\lambda_S$, $m_H$, $\lambda_H$ and $\lambda_{HS}$ are all free real parameters.

About the Lagrangian in Eq.(\ref{MDM}), one should note following points:
\begin{itemize}
\item
The MDM is actually a low energy theoretical framework describing the breakdown of a UV strong dynamics
with approximate scale invariance, and the dilaton in this theory is distinguished from the usual one.
Explicitly speaking, in the traditional dilaton models the whole SM sector is usually assumed to be
a part of the strong dynamics, and all the fermions and gauge bosons of the SM are composite particles
at the weak scale \cite{dilaton-before}. Under these theoretical assumptions,
the couplings of the linearized dilaton $S$ to the SM fields take following form \cite{dilaton-before}
\begin{eqnarray}
{\cal{L}} = \frac{S}{f} T^\mu_\mu,  \label{dilaton-SM}
\end{eqnarray}
where $T^\mu_\mu$ represents the trace of the energy-momentum tensor of the SM.
Through the interactions in Eq.(\ref{dilaton-SM}),
the dilaton couples directly to the fermions and $W$, $Z$ bosons in the SM
with the strengthes proportional to the mass of the involved particle.
In this way, the dilaton mimics the properties of the SM Higgs boson. By contrast, in the MDM
all SM particles except for the Higgs field are assumed to be the spectators of the strong dynamics,
and they are all elementary particles. As a result, the dilaton does not couple directly to these particles.

\item
In the original version of the MDM, the authors set $N_X = 1$
and chose the quantum numbers of the fermion $X_i$ same as those of the right-handed top quark.
This setting was motivated by topcolor theory \cite{Hill}, which intended to
present a reasonable explanation of the relatively large top quark mass within a minimal framework.
However, as we will show below, such a setting is tightly limited
by the vacuum stability of the theory at $m_{X_i}$ scale in interpreting the diphoton excess.
Considering that a strong dynamical theory usually involves rich fermion fields and the assignment on
their quantum numbers is somewhat arbitrary, we therefore consider a more general but also simple case,
which assumes that all the vector-like fermions are identical, and each of them transforms in the
$(3, 1, Y = 2 Q_X)$ representation of the SM gauge group
$SU(3)_c \bigotimes SU(2)_L \bigotimes U(1)_Y $. In the following, we vary the number of the fermions $N_X$,
their common mass $m_X$, and also their electric charge $Q_X$ to discuss the diphoton excess.
\end{itemize}

If one writes the Higgs field in unitary gauge via $\tilde{H} = \frac{1}{\sqrt{2}} U (0, H)^T$,  the scalar potential
in Eq.(\ref{potential}) can be rewritten as
\begin{eqnarray}
\tilde V(S,H)=\frac{m_S^2}{2} S^2+ {\lambda_S\over 4}S^4 + \frac{m_H^2}{2} H^2 + \frac{\lambda_H}{4} H^4 +
\frac{\lambda_{HS}}{4} S^2 H^2.  \label{potential1}
\end{eqnarray}
In the following, we consider the most general situation in which both $H$ and $S$ take vacuum expectation values (VEV),
$ \langle H \rangle = v $ and $\langle S \rangle = f$, and they mix to form mass eigenstates $h$ and $s$:
\begin{eqnarray}
  h &=& \cos\theta_S H + \sin\theta_S S, \nonumber\\
  s &=& - \sin\theta_S H +  \cos\theta_S S.
\end{eqnarray}
In our scheme for the diphoton excess, $h$ corresponds to the 125 GeV Higgs boson discovered at the LHC, and $s$ is responsible for the
$750 {\rm GeV}$ diphoton excess by the process $g g \to s \to \gamma \gamma$.
So in the following, we set $m_h=125 {\rm GeV}$, $m_s=750 {\rm GeV}$ and $v=246 {\rm GeV}$, and
for the convenience of our discussion, we choose $\eta \equiv \frac{v}{f} N_X$, $\sin \theta_S$, $Q_X$,
$N_X$ and $m_X$ as the input parameters of the MDM model.
In this case, we have following relations
\begin{eqnarray}
\lambda_{HS} &=& \frac{2 \eta (m_h^2 - m_s^2 ) \sin \theta_S \cos \theta_S}{v^2 N_X}, \nonumber \\
\lambda_H &=& \frac{m_h^2 \cos^2 \theta_S + m_s^2 \sin^2 \theta_S }{2 v^2}, \nonumber \\
\lambda_S &=& \frac{\eta^2 (m_h^2 \sin^2 \theta_S + m_s^2 \cos^2 \theta_S)}{2 v^2 N_X^2}. \label{parameters}
\end{eqnarray}
With the assumption that the dilaton is fully responsible for the fermion masses, the Yukawa coupling of
$X_i$ is given by $y_X \equiv \frac{m_X}{f} = \frac{\eta m_X}{v N_X}$. Obviously $y_X$
is inversely proportional to $N_X$ for fixed $\eta$ and $m_X$. As we will show below, the diphoton rate is only sensitive to
the parameters $\eta$, $\sin \theta_S$ and $Q_X$, and does not depend on $y_X$ directly.

\section{Useful formulae in getting the diphoton excess}

In the MDM, the particle $s$ may decay into $gg$, $\gamma \gamma$, $Z \gamma$, $ZZ$, $WW^\ast$, $f \bar{f}$ and $hh$.
In this section, we list the formulae for the widths of these decays, which are needed to get the diphoton rate.
As we will show below, these formulae are helpful to understand our results.

\begin{itemize}
\item The widths of $\phi \to \gamma \gamma, g g, Z \gamma$  with $\phi=h,s$:
\begin{eqnarray}
  \Gamma_{\phi\to \gamma\gamma} & = & \frac{G_\mu \alpha^2 m_\phi^3}{128\sqrt{2}\pi^3} \left|I_{\gamma}^\phi \right|^2, \label{loop1}\\
  \Gamma_{\phi\to gg} &=&  \frac{G_\mu \alpha_s^2 m_\phi^3}{16\sqrt{2}\pi^3} \left| I_{g}^\phi \right|^2,  \\
  \Gamma_{\phi\to Z\gamma}  &=&  \frac{G_\mu^2 m_W^2 \alpha m_\phi^3}{64\pi^4} \left( 1-\frac{m_Z^2}{m_\phi^2} \right)^3 \left| I_{Z \gamma}^\phi \right|^2,
\end{eqnarray}
where the $I_g^\phi$, $I_\gamma^\phi$ and $I^\phi_{Z\gamma}$ are given by
\begin{eqnarray}
I_{\gamma}^h &=& \cos \theta_S \times (A_1(\tau_W) + \frac{4}{3}A_{\frac{1}{2}}(\tau_t) ) +
\sin \theta_S N_c \eta Q_X^2 A_{\frac{1}{2}}(\tau_X),  \\
I_{\gamma}^s &=& - \sin \theta_S \times (A_1(\tau_W) + \frac{4}{3}A_{\frac{1}{2}}(\tau_t) ) +
\cos \theta_S N_c \eta Q_X^2 A_{\frac{1}{2}}(\tau_X), \label{diphoton-expression} \\
I_{g}^h &=& \frac{\cos \theta_S}{2} \times A_{\frac{1}{2}}(\tau_t)  +
\frac{\eta \sin \theta_S}{2} A_{\frac{1}{2}}(\tau_X),   \\
I_{g}^s &=& - \frac{\sin \theta_S}{2} \times A_{\frac{1}{2}}(\tau_t)  +
\frac{\eta \cos \theta_S}{2} A_{\frac{1}{2}}(\tau_X), 
\label{loop7}
\end{eqnarray}
\begin{eqnarray}
I_{Z\gamma}^h &=& \cos \theta_S \times ( \cos \theta_W C_1(\tau_W^{-1},\eta_W^{-1})
  +\frac{2(1-\frac{8}{3}\sin^2\theta_W)}{\cos \theta_W} C_{\frac{1}{2}}(\tau_t^{-1},\eta_t^{-1})) \nonumber \\
  & & +  4 \sin \theta_S N_c \eta Q_X^2 \frac{\sin^2 \theta_W}{\cos \theta_W} C_{\frac{1}{2}}(\tau_X^{-1},\eta_X^{-1}),  \\
I_{Z\gamma}^s &=& - \sin \theta_S \times ( \cos \theta_W C_1(\tau_W^{-1},\eta_W^{-1})
  +\frac{2(1-\frac{8}{3}\sin^2\theta_W)}{\cos \theta_W} C_{\frac{1}{2}}(\tau_t^{-1},\eta_t^{-1})) \nonumber \\
  & & +  4 \cos \theta_S N_c \eta Q_X^2 \frac{\sin^2 \theta_W}{\cos \theta_W} C_{\frac{1}{2}}(\tau_X^{-1},\eta_X^{-1}).
\label{width1}
\end{eqnarray}
In above expressions, $A_{\frac{1}{2}}$, $A_1$, $C_{\frac{1}{2}}$, $C_1$ are the loop functions defined in \cite{h-rev}
with $\tau_\beta =m_\phi^2/(4 m_{\beta}^{2})$ and $\eta_\beta = m_Z^2/(4 m_{\beta}^{2})$ for $\beta = W, t, X_i$.

About these formulae, one should note that the terms proportional to $\cos \theta_S$ in the expressions of $I_{i}^s$
are contributed by the dilaton component of $s$, while those proportional to $\sin \theta_S$ come from the
$H$-component of $s$. One should also note that in the case of $\sin \theta_S \sim 0$, which is required by
the null excess in the channels such as $ZZ$ and $hh$ at the $750 {\rm GeV}$ invariant mass (see below)
and also by the $125 {\rm GeV}$ Higgs data, $I_{\gamma}^s$, $I_{g}^s$ and $I_{Z\gamma}^s$ are all dominated
by the contribution from the vector-like fermions, and consequently they are correlated. Explicitly speaking, we have
$I_{\gamma}^s:I_{g}^s:I_{Z\gamma}^s = N_c Q_X^2 : \frac{1}{2}: \frac{N_c Q_X^2}{2} \frac{\sin^2
\theta_W}{\cos \theta_W}$ in the limit $m_s, m_X \gg m_Z$. This correlation may sever as a test of
the model at future LHC experiments.

\item The widths of the decays $s \to V V^\ast$ with $V = W, Z$.

If one parameterizes the effective $s V V^\ast$ interaction as
\begin{eqnarray}
{\cal{A}}_{s V V^\ast} = g_V m_V ( A_{V}^s g^{\mu \nu} + B_{V}^s p_2^{\mu} p_1^{\nu} ) \epsilon_\mu (p_1) \epsilon_{\nu} (p_2), \nonumber
\end{eqnarray}
then the decay width of $s \to V V^\ast$ is given by\cite{1512-COS}
\begin{eqnarray}
\Gamma_{s \to V V^\ast} &= & \delta_V
\frac{G_F m_s^3}{16 \pi \sqrt{2}} \frac{4 m_V^4}{m_s^4} \sqrt{\lambda(m_V^2,m_V^2;m_s^2)} \times \nonumber \\
&& \left[ A_{V}^s A^{s\ast}_{V} \times \left ( 2 + \frac{(p_1 \cdot p_2)^2}{m_V^4} \right ) + ( A_{V}^s B^{s\ast}_{V} + A^{s\ast}_{V} B_{V}^s) \times \left ( \frac{(p_1 \cdot p_2)^3}{m_V^4} - p_1 \cdot p_2 \right )  \right . \nonumber \\
&& \left. \ \ +\  B_{V}^s B^{s\ast}_{V} \times \left ( m_V^4 + \frac{(p_1 \cdot p_2)^4}{m_V^4} - 2 (p_1 \cdot p_2)^2 \right ) \right ], \label{Width-VV}
\end{eqnarray}
where $\delta_V=2(1)$ for $V=W(Z)$ respectively and $\lambda(x,y,z)= ((z-x-y)^2 - 4 xy)/z^2$.

In the MDM, we have
\begin{eqnarray}
A_W^s &\simeq& - \sin \theta_S, \quad B_W^s \simeq 0,    \nonumber \\
A_Z^s & \simeq & -\sin\theta_S + \frac{\alpha}{ 4 \pi m_Z^2} \cos \theta_S N_c \eta Q_X^2 \tan^2\theta_W p_1\cdot p_2 A_{\frac{1}{2}}(\tau_X), \nonumber \\
B_Z^s & \simeq & - \frac{\alpha}{ 4 \pi m_Z^2} \cos \theta_S N_c \eta Q_X^2 \tan^2\theta_W A_{\frac{1}{2}}(\tau_X). \nonumber
\end{eqnarray}

Note that in the expressions of $A_Z^s$ and $B_Z^s$, we have included the one-loop corrections. This is because in case
of $\sin \theta_S \sim 0$, the corrections are not always smaller than the tree level contributions. Also note that in
getting $A_Z^s$ and $B_Z^s$, to a good approximation we have neglected the $Z$ boson mass appeared in the loop functions,
and that is why we can express the corrections in term of the simple function $ A_{\frac{1}{2}}(\tau_X)$.

\item The width of the tree-level decay $s\to f\bar{f}$ with $f$ denoting any of the fermions in the SM:
\begin{eqnarray}
  \Gamma_{s\to f\bar{f}} &=& \sin^2\theta_S \frac{3G_\mu m^2_f m_s }{4\sqrt{2}\pi}
  \Big(1-\frac{4m_f^2}{m_s^2}\Big)^{\frac{3}{2}}.
\end{eqnarray}
Note that for this kind of decays, the widths are proportional to $\sin^2 \theta_S$.

\item The width of the tree level decay $s \to h h$:
\begin{eqnarray}
\Gamma_{s \to h h} &=& \frac{\left|C_{s hh}\right|^2}{16\pi m_{s}^2}\left(\frac{m_s^2}{4}-m_h^2\right)^\frac{1}{2}, \label{hh}
\end{eqnarray}
where
\begin{eqnarray}
C_{s h h} &=& - 6 \lambda_H v \sin\theta_S\cos^2\theta_S + 6 \lambda_S f \sin^2\theta_S\cos\theta_S \nonumber \\
          && + \lambda_{HS} (-v\sin^3\theta_S+f\cos^3\theta_S-2f\sin^2\theta_S \cos\theta_S+2v \sin\theta_S \cos^2\theta_S) \nonumber \\
          &\simeq& - \frac{2 m_s^2}{v} \sin \theta_S.  \nonumber
\end{eqnarray}
In getting the final expression of $C_{shh}$, we have used the relation $m_s^2 \gg m_h^2$
and $\sin \theta_S \sim 0$ to neglect some unimportant terms. Just like the decays
$s \to WW^\ast $ and $s \to t \bar{t}$, $\Gamma_{s \to h h}$ is proportional to $\sin^2 \theta_S$.
\end{itemize}

With these formulae, the total width of the scalar $s$ and the $s$-induced diphoton
rate can be written as
\begin{eqnarray}
\Gamma_{tot} &=& \Gamma_{s \to g g} + \Gamma_{s \to \gamma \gamma} + \Gamma_{s \to Z \gamma} + \Gamma_{s \to Z Z} + \Gamma_{s \to W W^\ast} + \Gamma_{s \to f \bar{f}} + \Gamma_{s \to h h} + \Gamma_{new},  \label{tot} \\
\sigma_{\gamma \gamma}^{13 TeV} &=& \frac{\Gamma_{\phi \to gg}}{\Gamma^{SM}_{H \to g g}} |_{m_H \simeq 750 {\rm GeV}} \times \sigma^{SM}_{\sqrt{s}=13 {\rm TeV}} (H) \times  \frac{\Gamma_{s\to\gamma\gamma}}{\Gamma_{tot}},
\end{eqnarray}
where the $\Gamma_{new}$ in Eq.(\ref{tot}) represents the contribution from the exotic decays of $s$, which may exist if the MDM is embedded
in a more complex theoretical framework, $\Gamma^{SM}_{H \to g g}$ denotes the decay width of the SM Higgs $H$ into $g g$ with $m_{H} = 750 {\rm GeV}$,
and $\sigma^{SM}_{\sqrt{s}=13 {\rm TeV}} (H)=735 {\rm fb}$  is the NNLO production rate of the $H$ at the 13 TeV LHC \cite{HCS}.  Obviously, if
$\Gamma_{tot}$ is determined mainly by $\Gamma_{gg}$, the rate can be approximated by
\begin{eqnarray}
\sigma_{\gamma \gamma}^{13 TeV} \simeq \frac{\Gamma_{\phi \to \gamma \gamma}}{\Gamma^{SM}_{H \to g g}} |_{m_H \simeq 750 {\rm GeV}} \times \sigma^{SM}_{\sqrt{s}=13 {\rm TeV}} (H)  \propto \eta^2 Q_X^4,
\end{eqnarray}
while if $\Gamma_{tot}$ takes a fixed value, we have
\begin{eqnarray}
\sigma_{\gamma \gamma}^{13 TeV} =  \left ( \frac{45 {\rm GeV}}{\Gamma_{tot}} \right ) \times \sigma_{norm} \times (\eta Q_X)^4,
\end{eqnarray}
where the normalized cross section $\sigma_{norm}$ is equal to $0.019 {\rm \ fb}$ ($0.018 {\rm \ fb}$) for $m_X = 1 {\rm TeV}$ ($1.5 {\rm TeV}$).

From the discussion in this section, one can get following important conclusions:
\begin{itemize}
\item The widths of $s \to gg, \gamma \gamma, Z \gamma$ or the production rates of the $gg$, $\gamma \gamma$ and $Z\gamma$ signals at the LHC
are correlated by
\begin{eqnarray}
\Gamma_{s \to g g} : \Gamma_{s \to \gamma \gamma} : \Gamma_{s \to Z \gamma} \simeq 1: \frac{9}{2} \frac{\alpha^2}{\alpha_s^2} Q_X^4: \frac{9}{4} \frac{\alpha^2}{\alpha_s^2} \tan^2 \theta_W Q_X^4 \simeq 1: 0.03 Q_X^4 : 0.004 Q_X^4.  \label{co-relation}
\end{eqnarray}
\item The widths listed from Eq.(\ref{loop1}) to Eq.(\ref{hh}) depend on the number of the vector-like fermions $N_X$ only through
the parameter $\eta \equiv \frac{v N_X}{f}$. As a result, explaining the diphoton excess puts non-trivial requirements on the combination $\frac{v N_X}{f}$, instead of
on the individual parameter $N_X$ or $y_X = \frac{\eta m_X}{v N_X}$.
\item Since the recent LHC searches for right-handed heavy quarks have required $m_X \gtrsim 900 \ {\rm GeV}$ \cite{1505-a-tp,1503-a-tp53,1601-c-tp53} and thus
$\tau_{X} \equiv m_s^2/(4 m_X^2) < 0.2$, the loop functions appeared in the widths change slightly with the further increase of $m_X$.
This implies that the widths and also the cross section have a very weak dependence
on the value of $m_X$. As a result, the results obtained in this work are only sensitive to the parameters $\eta$, $\sin \theta_S$ and $Q_X$.

At this stage, one can infer that  the parameter $N_X$ may also be understood as the total number of the vector-like fermions with the electric charge $Q_X$
in the strong dynamics because the contributions of the fermions to the diphoton rate are roughly identical. Since the particle content of a strong dynamics
is usually rich, $N_X$ is naturally larger than 1.
\end{itemize}

We remind that the second and third conclusions depend on the assumption that the dilaton is fully responsible for the masses of the vector-like fermions, and within our knowledge,
they were not paid attention to in previous literatures.

\section{Numerical results and discussions}

\begin{table}[t]
\caption{Upper limits on various $750 {\rm GeV}$ resonant signals at 8-TeV LHC set by either ATLAS or CMS collaboration \cite{1512-COS}. \label{tab1}}
\begin{tabular}{|c|c|c|c|c|c|c|}
  \hline
  Channel
  & $jj$ \cite{jj:ATLAS, jj:CMS} & $hh$ \cite{hh:ATLAS:1, hh:ATLAS:2, hh:ATLAS:3, hh:CMS} & $WW^\ast$ \cite{WW:ATLAS, VV:CMS} & $ZZ$ \cite{ZZ:ATLAS, VV:CMS}
   & Z$\gamma$ \cite{Zr:ATLAS} & $t\bar{t}$ \cite{tt:ATLAS, tt:CMS}  \\ \hline
  $95\%$ C.L. limits
  & 1800 {\rm fb} & 35 {\rm fb} & 37 {\rm fb} & 12 {\rm fb} & 3.6 {\rm fb} & 450 {\rm fb} \\
  \hline
\end{tabular}
\end{table}

In this section, we discuss the diphoton excess in the MDM. In order to get the favored parameter space for the excess,
we fix $Q_X=\frac{2}{3}$, $\frac{5}{3}$ and $m_X = 1 {\rm TeV}$, $1.5 {\rm TeV}$ at each time, and scan following parameter space
\begin{eqnarray}
0< \eta \leq 2, \quad \quad |\tan\theta_S| \leq 0.1.
\end{eqnarray}
During the scan, we consider following theoretical and experimental constraints:
\begin{itemize}
\item The vacuum stability at the scale of $m_s = 750 {\rm GeV}$ for the scalar potential,
 which corresponds to the requirement $4 \lambda_H \lambda_S - \lambda_{HS}^2 >0$ \cite{min-dilaton1}.
\item Constraints from the perturbativity at the scale of $m_s = 750 {\rm GeV}$, which requires
$\lambda_S, \lambda_H, \lambda_{HS} \lesssim 4 \pi$, and $y_X \lesssim 4 \pi/\sqrt{N_c}$ \cite{vacuum-3}.
\item Constraints from the electroweak precision data. We calculate the Peskin-Takeuchi $S$ and $T$
parameters \cite{STU} with the formulae presented in \cite{min-dilaton1}, and construct $\chi^2_{ST}$
by following experimental fit results with $m_{h,ref}=125\GeV$ and $m_{t,ref}=173\GeV$ \cite{Gfitter2014}:
\begin{eqnarray}
  S=0.06\pm0.09, ~~ T=0.10\pm0.07, ~~ \rho_{ST}=0.91.
\end{eqnarray}
In our calculation, we require that the samples satisfy $\chi^2_{ST} \leq 6.18$.
\item Experimental constraints from the 125 GeV Higgs data, which include the updated exclusive signal rates for
$\gamma \gamma$, $ZZ^\ast$, $W W^\ast$, $b\bar{b}$ and $\tau \bar{\tau}$ channels \cite{h-a-web, h-c-web}.
We perform the fits like our previous paper \cite{1311-MDM, 1309-LightHiggs},
and require the samples to coincide with the combined data at $2\sigma$ level.
\item Experimental constraints from the null results in the search for the 750 GeV resonance through other channels such as $s \to Z Z, hh$ at Run I,
just like what we did in \cite{1512-COS}. The upper bounds on these channels at $95\%$ C.L. are listed in Table \ref{tab1}.
\end{itemize}

\begin{figure}[t]
  \centering
\includegraphics[width=7.5cm]{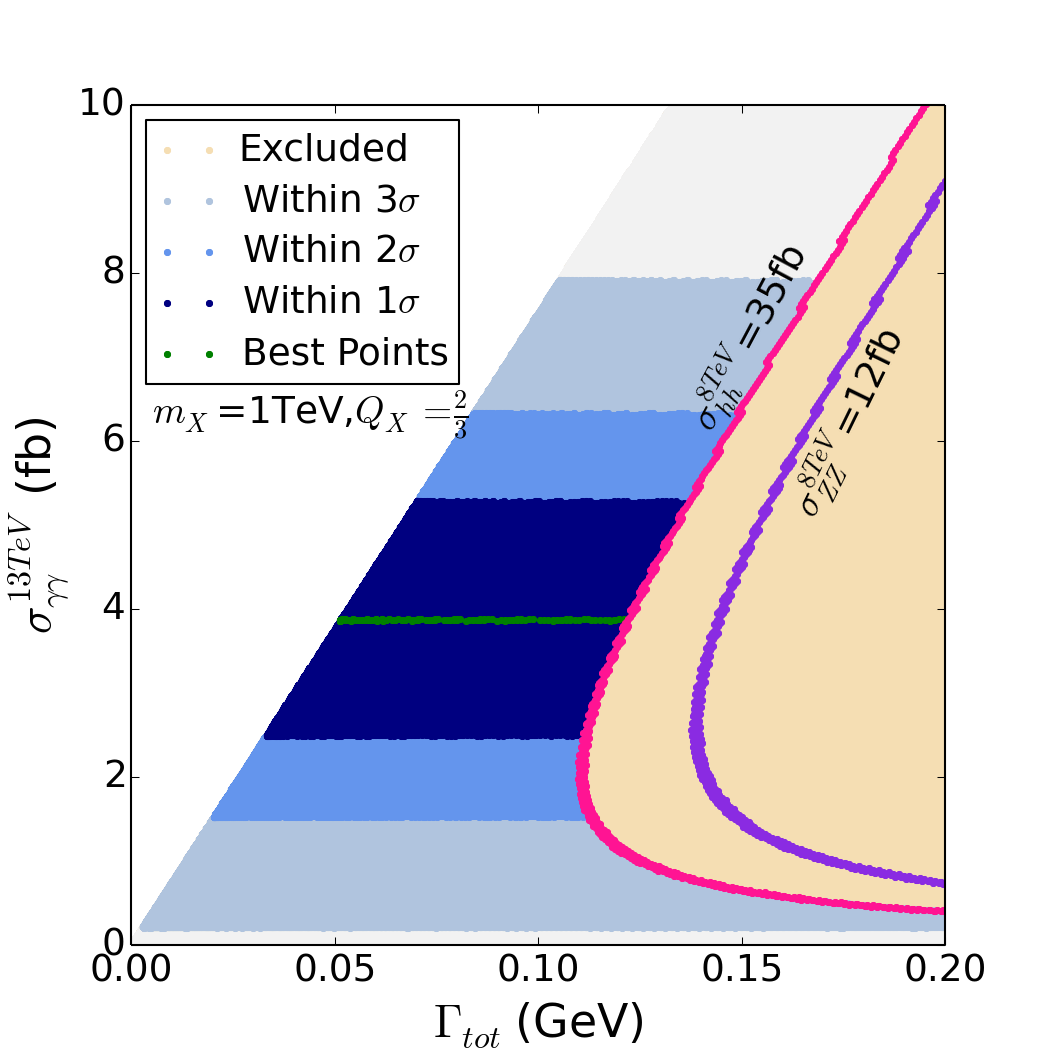}\hspace{-0.5cm}
\includegraphics[width=7.5cm]{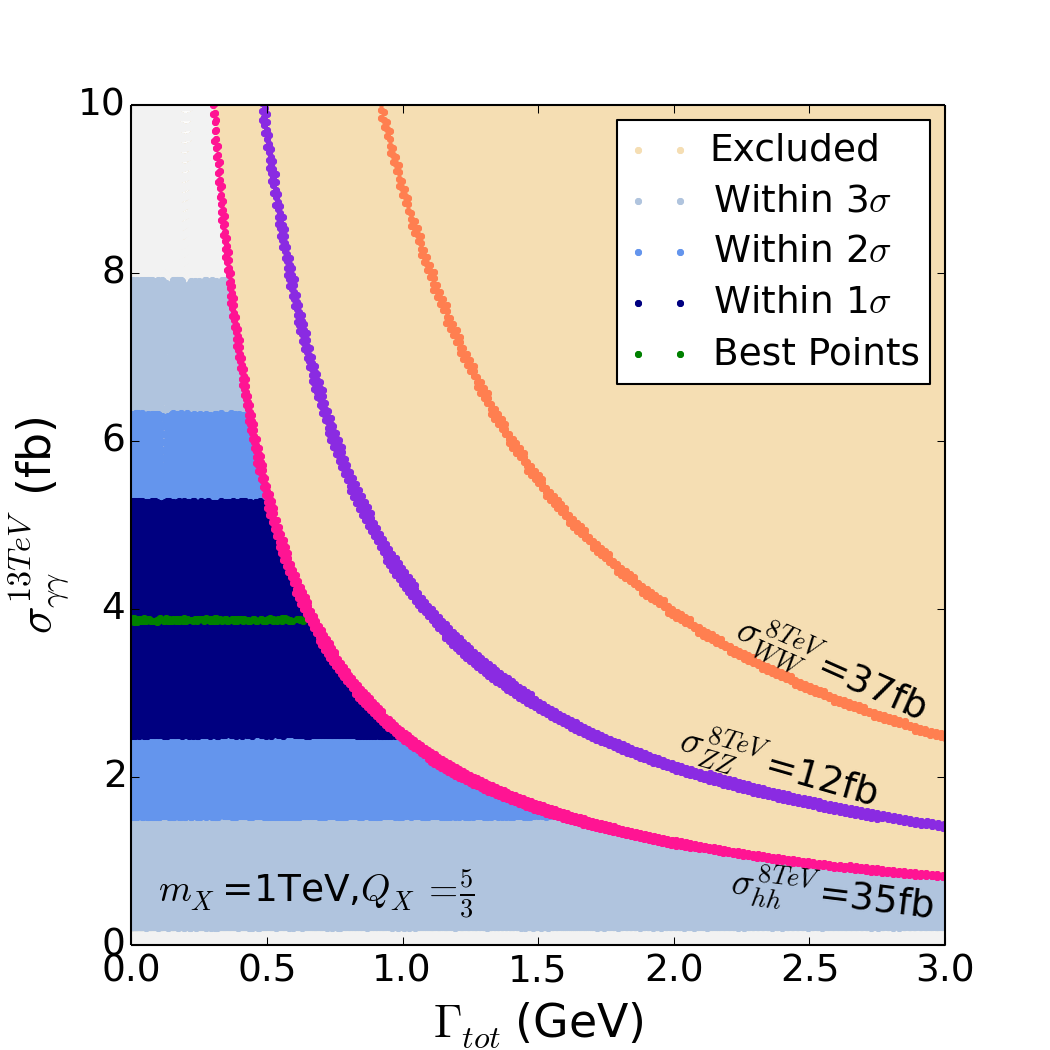}
\vspace*{-0.6cm}
\caption{The fit results of the MDM to the $750 {\rm GeV}$ diphoton data together with the LHC Run I constraints listed in Table \ref{tab1},
which are projected on the $\sigma_{\gamma \gamma}^{13 TeV}-\Gamma_{tot}$ planes for $Q_X=2/3$ (left panel) and  $Q_X=5/3$ (right panel)
respectively.  The regions filled by the colors from gray to deep blue represent the parameter spaces that can fit the diphoton data
within 3$\sigma$, 2$\sigma$ and 1$\sigma$ level respectively, and by contrast the regions covered by straw color are excluded by the
constraints.  The boundaries for the $hh$, $ZZ $ and $WW^\ast$ channels are also plotted, which correspond to blue lines, red lines
and brown lines respectively, and the other constraints listed in Table \ref{tab1} are too weak to be drawn on the panels. In each panel,
the green line represents the best-fit samples. In getting this panel, we have set $\Gamma_{new} = 0$ and  $m_X = 1 {\rm TeV}$, and we checked
that $m_X = 1.5 {\rm TeV}$  predicts roughly same results, which reflects that our results are insensitive to $m_X$.}
\label{fig1}
\end{figure}

For each sample surviving the constraints, we perform a fit to the $750 {\rm GeV}$ diphoton data collected at the $8\  {\rm TeV}$ and
the $13\  {\rm TeV}$ LHC. In doing this, we use the method introduced in \cite{Diphoton:UsedRate}, where the data were given by
\begin{eqnarray}\label{chisq}
\mu^{exp}_i & = &\sigma(pp\to \gamma\gamma)=\left\{\begin{array}{lllll}
0.63\pm 0.25 ~{\rm fb} 	& ~~~{\rm CMS}    &{\rm at} ~\sqrt{s} = 8 &~{\rm TeV }, \\
0.46\pm 0.85 ~{\rm fb} 	& ~~~{\rm ATLAS} &{\rm at} ~\sqrt{s} = 8 &~{\rm TeV},  \\
5.6\pm2.4   ~{\rm fb} 	& ~~~{\rm CMS}    &{\rm at} ~\sqrt{s} = 13 &~{\rm TeV }, \\
6.2^{+2.4}_{-2.0} ~{\rm fb}  	& ~~~{\rm ATLAS} &{\rm at} ~\sqrt{s} = 13 &~{\rm TeV}, \\
\end{array}\right.
\end{eqnarray}
and the $\chi^2_{\gamma \gamma}$ function was given by \cite{Diphoton:UsedRate, 1512-COS}
\begin{eqnarray}\label{chisq}
\chi^2 & = & \sum_{i=1}^4 \chi^2_i, \nonumber \\
\chi^2_i & = & \left\{\begin{array}{ll}
2[\mu_i^{exp} - \mu_i + \mu_i {\rm ln} \frac{\mu_i}{\mu_i^{exp}} ] 	& ~~~{\rm for~ the~ 13~TeV~ATLAS~ data}, \\
\frac{(\mu_i^{exp}-\mu_i)^2}{\sigma_{\mu_i^{exp}}^2}	& ~~~{\rm for~the~other~three~sets~of~data}, \\
\end{array}\right.
\end{eqnarray}
with $\mu_i$ denoting the theoretical prediction of the diphoton rate.

\begin{figure}[t]
  \centering
\includegraphics[width=7.5cm]{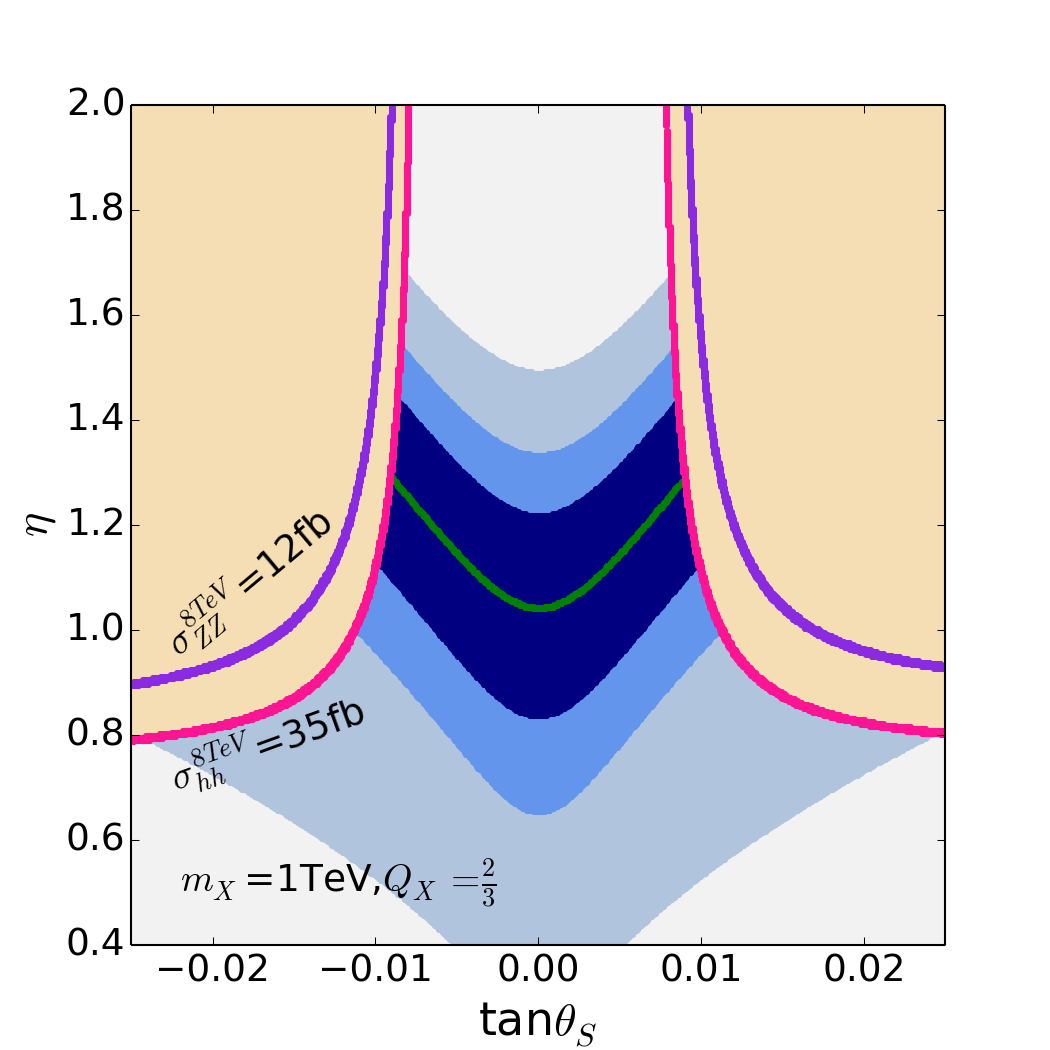}\hspace{-0.5cm}
\includegraphics[width=7.5cm]{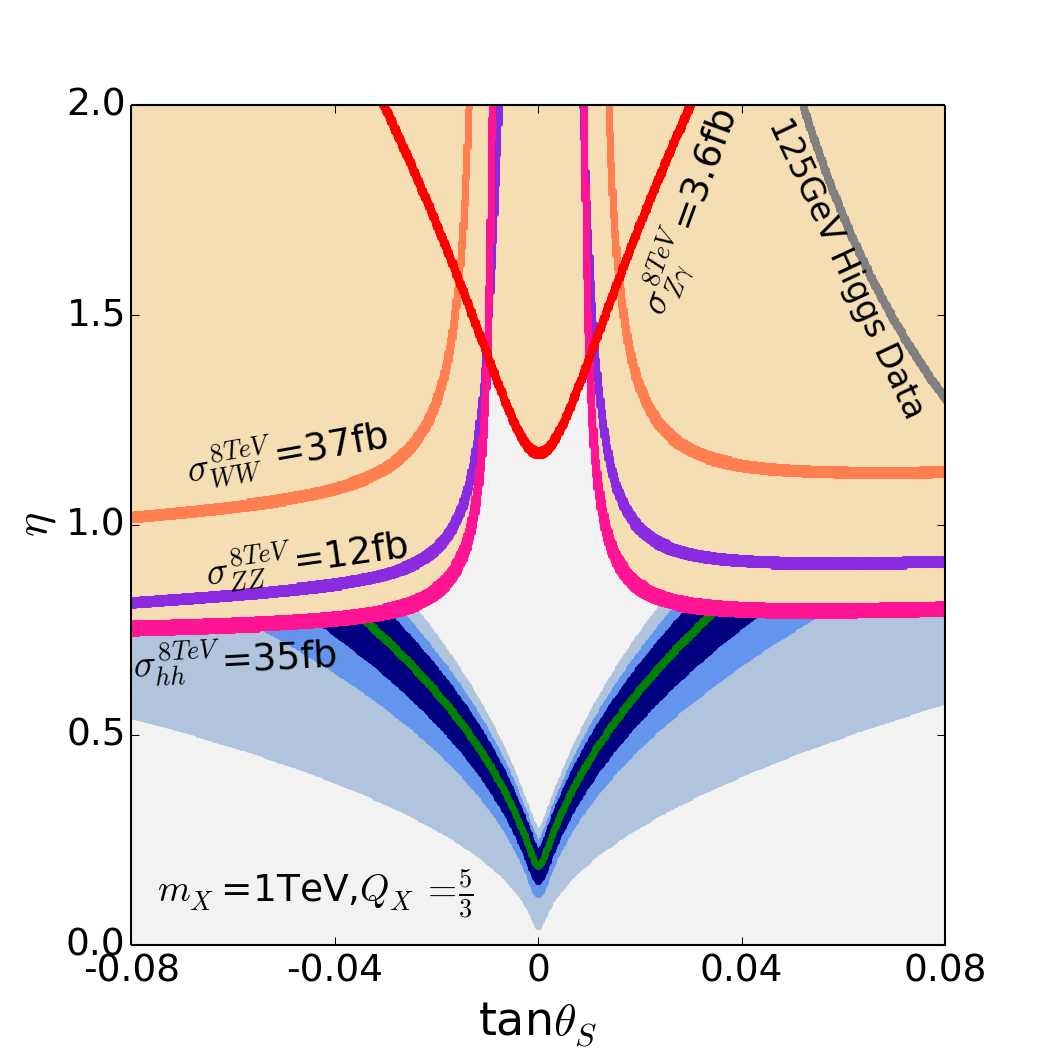}
\vspace*{-0.6cm}
\caption{Same samples as those in Fig.\ref{fig1}, but projected on the $\eta-\tan\theta_S$ planes. Although we take  $m_X = 1 {\rm TeV}$
in getting this figure,  we check that setting $m_X = 1.5 {\rm TeV}$ produces indistinguishable difference on the figure due to the
comments below Eq.(\ref{co-relation}). }
\label{fig2}
\end{figure}

In the following, we only consider the samples surviving the first four constraints.  In Fig.\ref{fig1}, we project these samples on
the $\sigma_{\gamma \gamma}^{13 TeV}-\Gamma_{tot}$ planes for $Q_X=2/3$ (left panel) and  $Q_X=5/3$ (right panel) respectively.
The details of this figure are explained in its caption. From this figure, one can get following facts:
\begin{itemize}
\item The central value of the diphoton rate is $3.9 {\rm fb}$ at the $13 {\rm TeV}$ LHC from the fit, and the $1\sigma$, $2 \sigma$ and
$3 \sigma$ ranges of the rate are $(2.5 \sim 5.3)  {\rm \ fb}$, $(1.5 \sim 6.3) {\rm \  fb}$, $ (0.2 \sim 7.9) {\rm \ fb}$
respectively. Note that this conclusion is independent of the value of $Q_X$.
\item For both $Q_X= \frac{2}{3}$ and $Q_X= \frac{5}{3}$ cases, the diphoton excess can be well explained. The difference of the two options
comes from the fact that for $Q_X= \frac{2}{3}$ case, $\Gamma_{tot} \lesssim 0.15 {\rm GeV} $ if one wants to explain the excess at $2\sigma$ level, while for
$Q_X= \frac{5}{3}$ case, $\Gamma_{tot} \lesssim 1.6 {\rm GeV} $. The reason for such a difference is that in the $Q_X= \frac{5}{3}$  case,
$\sin \theta_S$ can take a larger value (see discussion below).
\item Among the channels listed in Table \ref{tab1}, the $hh$ channel puts the tightest constraints on the parameter space regardless the value of $Q_X$.
\end{itemize}

Next we illustrate the favored parameter regions for the excess. For this purpose, we project the samples used in Fig.\ref{fig1} on the
$\eta-\tan\theta_S$ planes, which are shown in Fig.\ref{fig2}.  This figure indicates following facts:
\begin{itemize}
\item In order to explain the diphoton excess at $2 \sigma$ level, $ 0.65 \leq \eta \leq 1.55$ and
$|\tan\theta_S| \leq 0.012$ are preferred for $Q_X = \frac{2}{3}$ case, and by contrast $ 0.15 \leq \eta \leq 0.8$ and
$|\tan\theta_S| \leq 0.06$ are preferred for $Q_X = \frac{5}{3}$ case. Note that in the  $Q_X = \frac{5}{3}$ case, a smaller $\eta$
as well as a wider range of $\tan \theta_S$ are favored to explain the excess in comparison with the  $Q_X = \frac{2}{3}$ case. The reason is that
a larger $Q_X$ can increase greatly the width and also the branching ratio of $s \to \gamma \gamma$, which
in return needs a smaller $s$ production rate to explain the excess.
\item The channels listed in Table \ref{tab1} exclude the parameter space characterized by a large $\eta$ and/or a large
$|\tan \theta_S|$. For these cases, the production rates of the channels are usually enhanced, which
can be inferred from the expressions of the widths.
\item In case of $\tan \theta_S \simeq 0$, the $Z \gamma$ channel may impose upper bounds on $\eta$, which is shown in the right panel of Fig.\ref{fig2}.
\item The favored parameter space is not symmetric if the sign for $\tan \theta_S$ is reversed, and this asymmetry turns out to be more obvious for
larger $Q_X$ and $|\tan \theta_S|$. The source of such a asymmetry comes from the expressions of $\Gamma_{s \to g g}$, $\Gamma_{s \to \gamma \gamma}$,
$\Gamma_{s \to Z \gamma}$ and $\Gamma_{s \to Z Z}$, which are presented from Eq.(\ref{loop1}) to Eq.(\ref{Width-VV}).
\end{itemize}

\begin{table}[t]
\small
\centering
\caption{Detailed information for one of the best points in the left and right panels of Fig.\ref{fig2} (labeled by P1 and P2 hereafter) respectively.
We checked that all these points predict $\chi^2_{\gamma\gamma}=2.32$, which corresponds to a $p$-value of 0.68. }
\label{tab2}
\begin{tabular}{ccccccccccccccc}
\hline
\hline
Point ~&~ $Q_{X}$     ~&~ $\eta$  ~&~  tan$\theta_S$ ~&~  $\frac{\Gamma_{\phi\to gg}}{\Gamma_{H\to gg}^{SM}}$ ~&~ BR$_{\phi\to g g}$  ~&~ BR$_{\phi\to \gamma\gamma}$  ~&~ BR$_{\phi\to ZZ}$ ~&~ BR$_{\phi\to WW^\ast}$ & BR$_{\phi\to hh}$ ~&~  BR$_{\phi\to t\bar{t}}$\\
\hline
$P_1$ ~&~ $\frac{2}{3}$ ~&~  1.144 ~&~ -0.005 ~&~ 0.973 ~&~ 82.1\% ~&~ 0.54\% ~&~ 2.4\% ~&~ 4.85\%  ~&~ 9.00\% ~&~ 1.02\% \\
$P_2$ ~&~ $\frac{5}{3}$ ~&~  0.336 ~&~ -0.005 ~&~ 0.083 ~&~ 24.4\% ~&~ 6.34\% ~&~ 9.62\% ~&~ 19.23\% ~&~ 35.60\% ~&~4.05\% \\
\hline
\end{tabular}
\end{table}

In Table \ref{tab2}, we show the detailed information for one of the best points in the left and right panels of Fig.\ref{fig2} respectively. In the following,
we label the two points by $P_1$ and $P_2$ respectively. From this table, one can learn that to explain the diphoton excess in the MDM, the branching ratio of
$s \to \gamma \gamma$ is usually at $1\%$ level, which is significantly larger than that of the Higgs boson in the SM.
One can also learn that for the best points, $s \to g g$ may be either dominant or subdominant decay channel of the $s$.


\begin{figure}[t]
  \centering
\includegraphics[width=14cm,height=11cm]{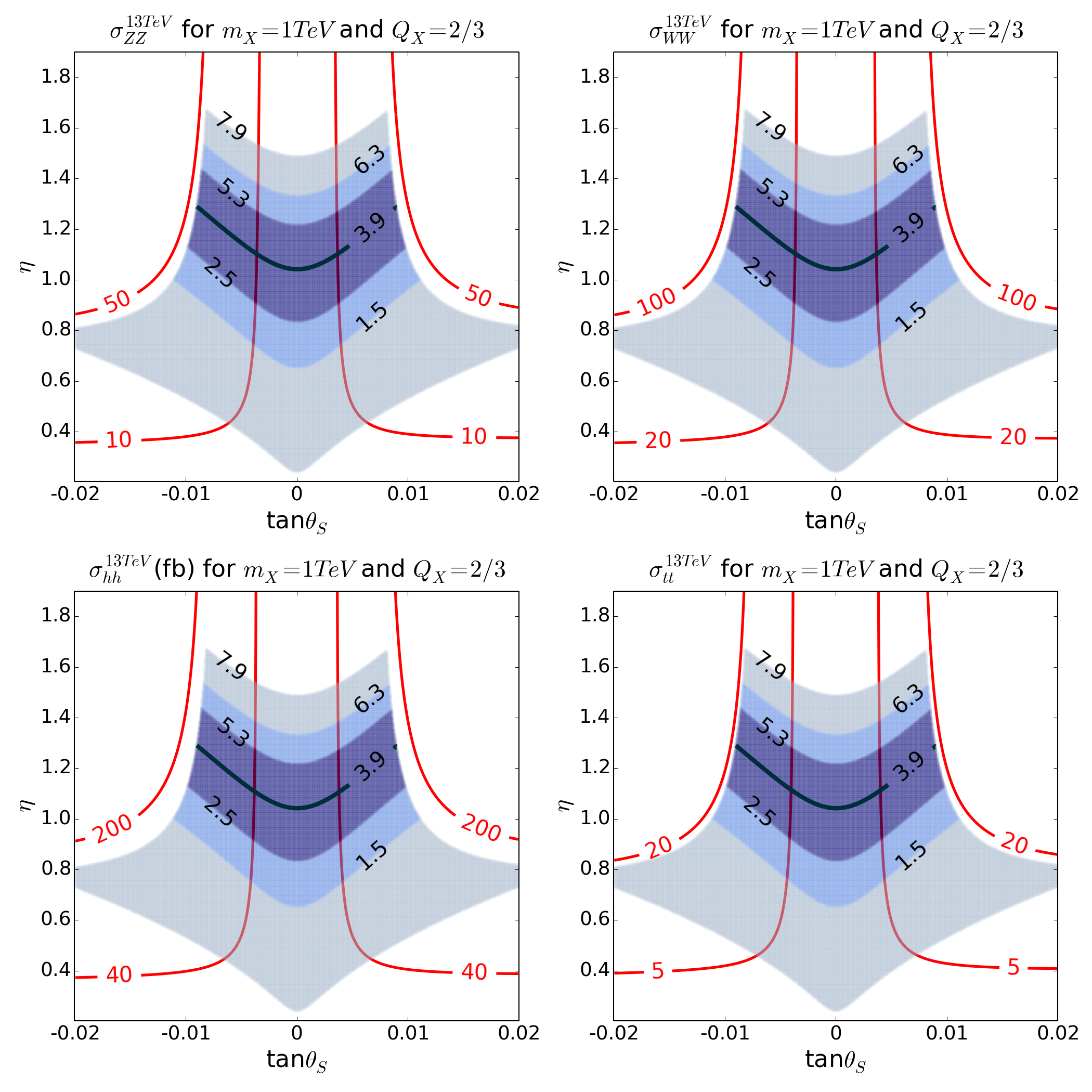}
\vspace*{-0.6cm}
\caption{Correlations of the diphoton rate at the $13 {\rm TeV}$ with those of $ZZ$, $WW^\ast$, $hh$ and
$t\bar{t}$ signals respectively for the $Q_X = \frac{2}{3}$ case, which are shown on the $\eta-\tan\theta_S$ planes.
Colors in this figure have same meanings as those in Fig.\ref{fig2}, and from the left to right and upper
to lower panels, the constant contours (red lines) of the production rates for $ZZ$, $WW^\ast$, $hh$ and $t\bar{t}$
signals are shown respectively. The numbers on the red lines represent the corresponding
production rates at the $13 {\rm TeV}$ LHC.  Note that the correlations of the diphoton rate
with those of the $gg$ and $Z\gamma$ signals are presented in Eq.(\ref{co-relation}).}
\label{fig4}
\end{figure}

\begin{figure}[t]
  \centering
\includegraphics[width=14cm,height=11cm]{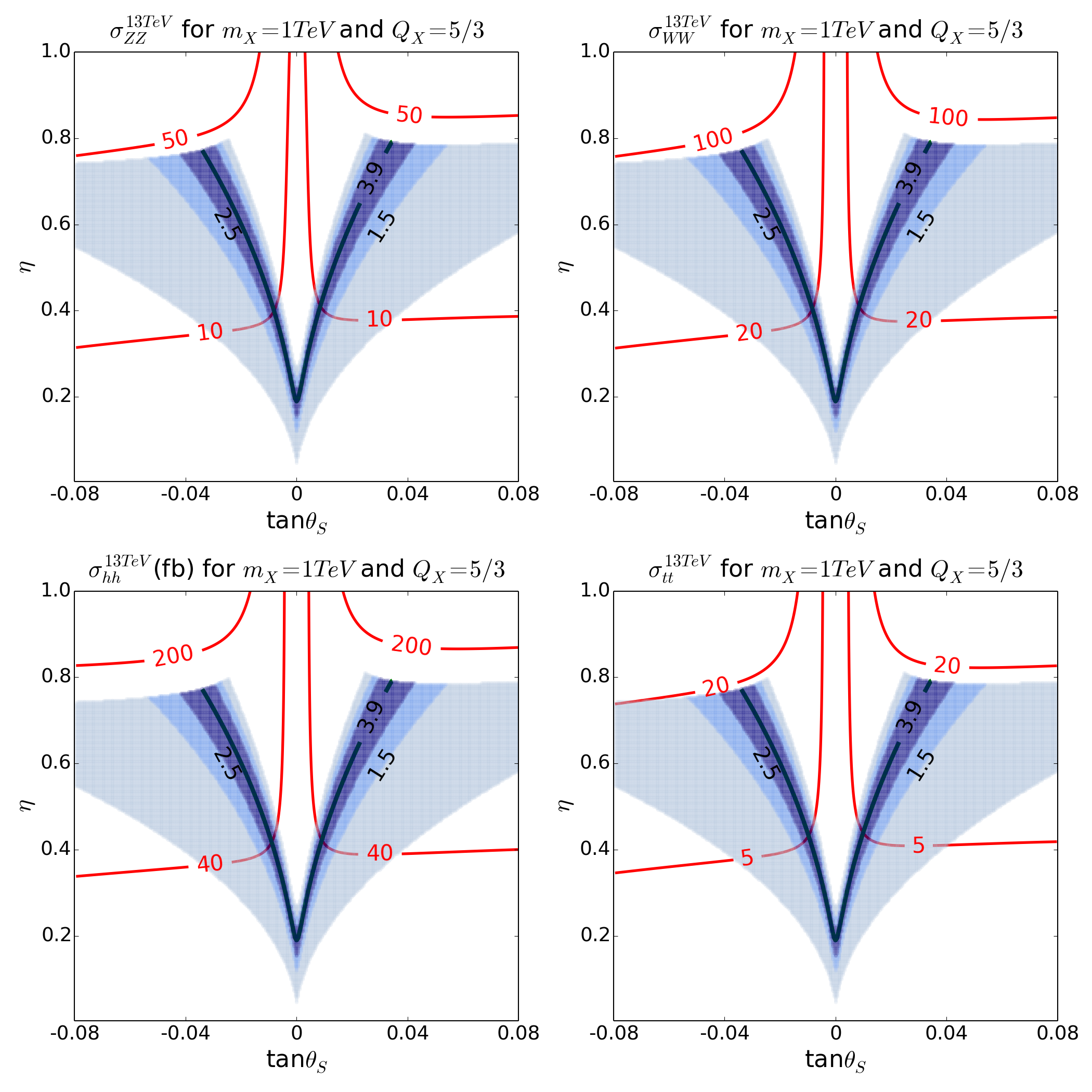}
\vspace*{-0.6cm}
\caption{Similar to Fig.\ref{fig4}, but for the $Q_X = \frac{5}{3}$ case.}
\label{fig5}
\end{figure}

Finally, we study the correlations between the diphoton rate at the $13 {\rm TeV}$ LHC with the rates of
the $ZZ$, $WW^\ast$, $hh$ and $t\bar{t}$ signals respectively. The results are presented in Fig.\ref{fig4} for the
$Q_X = \frac{2}{3}$ case with the implication of the figure explained in its caption. This figure reveals
following information
\begin{itemize}
\item Current LHC data have put upper limits on the rates of the different signals at the $13-{\rm TeV}$ LHC, which are
$\sigma_{ZZ} \lesssim 48 {\rm fb}$,  $\sigma_{WW} \lesssim 96 {\rm fb}$, $\sigma_{hh} \lesssim 190 {\rm fb}$
and $\sigma_{t\bar{t}} \lesssim 19 {\rm fb}$.
\item Since for a moderately small $\sin\theta_S$, the $sZZ$, $sWW$, $shh$ and $st\bar{t}$ couplings are roughly
proportional to $\sin\theta_S\simeq\tan \theta_S$, the constant contours of the signal rates exhibit similar behaviors on the
$\eta-\tan\theta_S$ plane. Obviously, if the diphoton excess persists at future LHC experiments
and meanwhile none of the other signals is observed, a small $\tan \theta_S$ is preferred.
\item More important, if more than one type of the signals are measured at the future LHC experiments,
one can decide the parameters of the MDM.
For example, given that $\sigma_{\gamma \gamma}$ and $\sigma_{jj}$ are precisely known,
one can get the value of $Q_X$, and if $\sigma_{\gamma \gamma}$ and $\sigma_{ZZ}$ are also
measured, one can pin down the favored regions of $\eta$ and $\sin \theta_S$.
\end{itemize}

In Fig.\ref{fig5}, we show the correlations of the different signals for the $Q_X = \frac{5}{3}$ case. The features of this figure
are quite similar to those of Fig.\ref{fig4} except that:
i) now the diphoton rate becomes more sensitive to $\eta$ and $\sin \theta_S$,
So to extract the values of the two parameters in this case, a more precise measurement of the diphoton signal is needed.
ii) the asymmetry between $\pm\tan\theta_S$ on the rates at 13TeV LHC becomes more obvious.

\begin{table}[t]
\centering
\caption{The scale where the vacuum becomes unstable for different choice of
the vector-like fermion number $N_X$.  Here the scale $\mu$ is in unit of GeV, and the
points $P_1$ and $P_2$ correspond to the two benchmark points in Table \ref{tab2}.
We checked that for the point P2 with $N_X =5, 6$, the vacuum keeps stable before $\lambda_H$ reaches
its Landau poles, which are roughly at $5.6 \times 10^{11} {\rm GeV}$ and $3.8 \times 10^{10} {\rm GeV}$
respectively. We also checked that for the $P_2$ with $N_X = 4 $, the Landau pole of $\lambda_H$
is roughly at $2.2 \times 10^{13} {\rm GeV}$.  \label{tab3}}

\vspace{0.3cm}

\begin{tabular}{|c|c|c|c|c|c|c|}
\hline
 Point   & $\mu (N_X = 1)$  & $\mu (N_X = 2)$  &$\mu (N_X = 3)$  &$\mu (N_X = 4)$  &$\mu (N_X = 5)$  &$\mu (N_X = 6)$\\
 \hline
$P_1$ & 1000  & 1000  & 1000  & 1015   & 1180  &1350 \\
\hline
$P_2$ & 1000 & 4930 & 57950 &  $2.1\times 10^7$ &$-$ &$-$\\
\hline
\end{tabular}
\end{table}

\section{Vacuum stability at high energy scale}

About one week before we finish this work, several papers appeared to discuss the vacuum stability
in a theoretical framework which is quite similar to the MDM \cite{vacuum-1,vacuum-2,vacuum-3}. The main argument of these papers was that,
in order to explain  the diphoton excess, the Yukawa coupling $y_X$ must be so large that the vacuum
becomes unstable at a certain high energy scale \footnote{The large Yukawa coupling $Y_X$ can influence the vacuum
stability condition $4 \lambda_H \lambda_S - \lambda_{HS}^2 > 0 $ by two ways \cite{vacuum-3}.
One is that it pulls down the value of $\lambda_S$ in its evolution with the energy scale by the renormalization
group equation (RGE). The other is that the threshold correction to the $\lambda_S$ at the scale $m_X$ is proportional to
$- y_X^4$, and consequently $\lambda_S$ usually becomes negative after considering the correction. }.
In our opinion, the MDM may be free of this problem due to following two reasons. One is that the MDM
is actually a low energy effective theory describing the breakdown of a
strong dynamics with approximate scale invariance. This means that the physics beyond the MDM must appear at a certain
high energy scale. The other is that, as we emphasized in Section III, the diphoton excess actually imposes
non-trivial requirements on the parameter $\eta \equiv \frac{v N_X}{f}$, instead of on the Yukawa coupling
$y_X \equiv \frac{\eta m_X}{v N_X}$ directly. For a given value of $\eta$, one may increase $N_X$ to suppress
the Yukawa coupling $y_X$, and thus alleviate the problem. In order to verify our speculation, we assume that there
are no particles in the strong interaction sector other than the vector-like fermions, and consider
the two benchmark points presented in Table \ref{tab2}. We repeat the analysis in \cite{vacuum-3}, i.e. we use
the same RGEs as those in \cite{vacuum-3} to run all parameters in the MDM, and also consider the threshold
correction to $\lambda_S$ at the scale $m_X$.  In Table \ref{tab3}, we present the scale where the vacuum
becomes unstable for different choices of $N_X$. This table indicates that moderately large $N_X$ and $Q_X$
are helpful to stabilize the vacuum state.

Finally, we remind that, although large $Q_X$ and/or $N_X$ are welcomed to explain the excess, they can not
be arbitrarily large in the extension of the SM by one gauge singlet scalar and the vector-like fermions.
The reason is that the $\beta$ function of the gauge coupling
$g_1$ is given by $\beta_{g_1} = ( \frac{41}{10} + N_X Q_X^2 \frac{12}{5} ) g_1^3$  \cite{vacuum-3},
and consequently $g_1$ increases rapidly with the RGE energy scale for large $N_X$ and $Q_X$.
In this case, the $\beta$ function of $\lambda_H$ is dominated by the term proportional to $g_1^4$, and
consequently, $\lambda_H$ may reach its Landau pole at an energy scale not far above the weak scale.

\section{conclusion}

The MDM extends the SM by adding vector-like fermions and one gauge singlet scalar, which represents a linearized dilaton field.
In this theory, the couplings of the dilaton to $gg$ and $\gamma \gamma$ are induced by the loops of the vector-like fermions, and may
be sizable in comparison with the $H g g$ and $H \gamma \gamma$ couplings in the SM. On the other hand, due to the singlet nature of the
dilaton its decays into the other SM particles are suppressed. These characters make the diphoton signal of the dilaton potentially
detectable at the LHC.

In this work, we tried to interpret the diphoton excess recently reported by the ATLAS and CMS collaborations
at the $13 {\rm\  TeV}$ LHC in the framework of the MDM. For this purpose, we first showed by analytic formulae
that the production rates of the $\gamma \gamma$, $gg$, $Z\gamma$, $ZZ$,
$WW^\ast$, $t\bar{t}$ and $hh$ signals at the $750 {\rm GeV}$ resonance are only sensitive to
the dilaton-Higgs mixing angle $\theta_S$ and the parameter $\eta \equiv v N_X/f$,
where  $N_X$ denotes  the number of the vector-like fermions and $f$ is the dilaton decay constant.
Then we scanned the two parameters to find the solutions to the excess. During the scan, we considered
various theoretical and experimental constraints, which included the vacuum stability and the perturbativity
of the theory at the scale of $m_s$, the electroweak precision data,  the $125 {\rm GeV}$
Higgs data, the LHC searches for exotic quarks, and the upper bounds on the rates of $ZZ$, $WW^\ast$,
$Z\gamma$, $t\bar{t}$ and $hh$ signals at LHC Run I.
We concluded that the model can predict the central value of the diphoton rate without conflicting
with any constraints. Moreover, after deciding the parameter space for the excess we discussed the
signatures of the theory at the LHC Run II. We showed that the rates of the $WW^\ast$ and $hh$ signals
may still reach about $100 {\rm \ fb}$ and $200 {\rm \ fb}$ respectively at the $13 {\rm \ TeV}$ LHC, and
thus they provide good prospect for detection in future.

As an indispensable part of this work, we also discussed the vacuum stability of the theory at high energy scales.
We showed that, by choosing moderately large $N_X$ and $Q_X$, the vacuum in our explanation can retain stable
up to $10^{11} {\rm GeV}$.

{\bf{Note added:}} When we finished this work at the beginning of this January, we noted that two papers had
appeared trying to explain the diphoton excess with the dilaton field \cite{Dilaton-1,Dilaton-2}. However, after
reading these papers, we learned that the paper \cite{Dilaton-1} considered the traditional dilaton model,
and the paper \cite{Dilaton-2} focused on 5D warped models. So their studies are quite different from ours.
We also noted that by then there existed several papers studying the diphoton excess in the model which extends the SM
by one gauge singlet scalar field and vector-like fermions \cite{general-singlet-extension,vacuum-1,vacuum-3}.
Compared with these works, our study has following features (improvements):
\begin{itemize}
\item We considered a generic model which predicts $N_X$ vector-like fermions (by contrast, most of the previous studies
considered the most economical $N_X = 1$ case). This enables us to explain the
diphoton excess without invoking a large Yukawa coupling $y_X$. Such a treatment, as we have discussed in section V,
is helpful to retain vacuum stability of the theory at high energy scales.
\item More important, by assuming that the dilaton field is fully responsible for the masses of the vector-like
fermions, we showed by analytic formulae that the rates for all the signals discussed in this work, such as $\gamma \gamma$,
$gg$, $Z \gamma$, $VV^\ast$, $f\bar{f}$ and $hh$, are only sensitive to the parameter $\eta = \frac{v N_X}{f}$, the dilaton-Higgs
mixing angle $\theta_S$ and the electric charge of the fermions $Q_X$. This observation can greatly simplify the
analysis on the diphoton excess, and within our knowledge, it was not paid due attention in previous studies.
\item We considered various constraints on the model, especially those from different observations at the LHC Run I
(which were listed in Table \ref{tab1}), and we concluded that the $hh$ signal usually puts the tightest constraint on our
explanation. This conclusion is rather new. Moreover, we also studied the signatures of our explanation at the LHC Run II,
which are helpful to decide the parameters of the model. Such a study was absent in previous literatures.
\end{itemize}

Before we end this work, we'd like to clarify its relation with our previous work \cite{1512-COS},
where we utilized the singlet extension of the Manohar-Wise model to explain the diphoton excess.
In either of the works, the scalar sector of the considered model contains a doublet and a singlet scalar field,
which mix to form a 125 GeV SM-like Higgs $h$ and a 750 GeV new scalar $s$, and the $s\gamma \gamma$ and $s g g$
interactions are induced by colored particles through loop effects. In organizing these works,
we first introduced the theoretical framework and listed the formula for the partial
widths of the scalar $s$, then we analyzed various constraints on the model
and discussed the diphoton signal from the process $gg \to s \to \gamma \gamma$. We concluded
that both the models can predict the central value of the excess in their vast parameter space.  Since the
two works adopted same $\chi^2$ function for the excess which only depends on
the diphoton rate, the $\chi^2$ values for the best points are same in the two explanations.
In spite of these similarities, we still think that the two works are independent since they are based
on different physics. The differences are reflected in following aspects:
\begin{itemize}
\item The origin of the singlet dominated scalar $s$. In the work \cite{1512-COS}, the singlet field is imposed by hand and only
for interpreting the excess, while in this work it corresponds to a linearized dilation field, which is well motivated by the broken of a strong
dynamic with approximate scale invariance.

\item The mechanism to generate sizable $s\gamma \gamma$ and $s g g$ interactions. In the singlet extension of the Manohar-Wise model, these
interactions are induced by color-octet and isospin-doublet scalars $S_R^A$, $S_I^A$ and $S_\pm^A$ with $A=1, \cdots 8$ denoting color index
(Note that there are totally $32$ bosonic freedom), so their coupling strengths are proportional to $(C_{s S_i^{A \ast} S_i^A} v)/m_{S_i}^2 A_0(\tau_{S_i})$
with $C_{s S_i^{A \ast} S_i^A}$ denoting the coupling coefficient for the $s S_i^{A \ast} S_i^A$ interaction. As a comparison, the couplings in this work
are induced by the vector-like fermions, and their strengthes are determined by the factor $\eta A_{\frac{1}{2}} (\tau_X)$.  Since the loop function $A_0$
is usually several times smaller than the function $A_{\frac{1}{2}}$\cite{h-rev}, beside the large bosonic freedom, large $C_{s S_i^{A \ast} S_i^A}$ and
meanwhile moderately light $S_i^A$ are also necessary to get the same sizes of the strengthes as those in this work.
By contrast,  we only need to tune the value $\eta$ to get the right couplings for the excess in this work. So the explanation presented in here
is rather simple and straightforward.

\item The intrinsic features of the explanations. Due to the particle assignments of the models, the two explanations exhibit
different features. For example, for the explanation in \cite{1512-COS} the upper limit of the dijet channel
in Table \ref{tab1} has constrained the diphoton rate to be less than about $7.5 {\rm fb}$ \cite{1512-COS,Staub}, while in the present
work the constraint from the dijet channel on the rate is rather loose. Another example is that for the explanation in \cite{1512-COS},
the vacuum stability can never constrain the model parameters, while
in this work it acts as a main motivation to consider moderately large $N_X$ and $Q_X$ to keep the vacuum stability.

\end{itemize}

\section*{Acknowledgement}
We thank Prof. C. P. Yuan and Fei Wang for helpful discussion, and this work was supported in part by the National Natural
Science Foundation of China (NNSFC) under Grant No. 11547103, 11275245, 11547310, 11575053.
Dr. Zhu thanks the support of the U.S. National Science Foundation under
Grant No. PHY-0855561, while he was working at Michigan State University.

\end{document}